\documentclass[twocolumn,showpacs,preprintnumbers,amsmath,amssymb,prb,superscriptaddress,aps,10pt]{revtex4-1}

\usepackage{graphicx}
\usepackage{color}

\begin{document}

\title{Transient dynamics and steady state behavior of the Anderson-Holstein model with a superconducting lead}
\author{K. F. Albrecht}
\altaffiliation{These authors contributed equally to this work.}
\affiliation{Physikalisches Institut, Universit\"at Freiburg, Hermann-Herder-Str. 3, D-79104 Freiburg, Germany}
\author{H. Soller}
\altaffiliation{These authors contributed equally to this work.}
\affiliation{Institut f\"ur Theoretische Physik, Ruprecht-Karls-Universit\"at Heidelberg, Philosophenweg 19, D-69120 Heidelberg, Germany}
\author{L. M\"uhlbacher}
\affiliation{Physikalisches Institut, Universit\"at Freiburg, Hermann-Herder-Str. 3, D-79104 Freiburg, Germany}
\author{A. Komnik}
\affiliation{Institut f\"ur Theoretische Physik, Ruprecht-Karls-Universit\"at Heidelberg, Philosophenweg 19, D-69120 Heidelberg, Germany}

\date{\today}

\begin{abstract}
  We analyze the nonequilibrium dynamics and steady-state behavior of the two-terminal Anderson-Holstein model with a superconducting and a normal conducting lead. In the deep Kondo limit we develop an analytical description if no phonons are included and a rate equation approach when phonons are present. Both cases are compared with the numerically exact diagrammatic Monte Carlo method obtaining a good agreement. For small voltages we find a pronounced enhancement of phonon sidebands due to the SC DOS.
\end{abstract}

\pacs{
  74.45.+c,
  73.63.Kv,
  5.10.Ln,
  72.10.Di
}

\maketitle

\section{Introduction}

In recent years there has been considerable interest in the study of quantum transport phenomena in superconductor (SC) hybrid structures driven out of equilibrium.\cite{Mourik12042012,2009Natur.461..960H,PhysRevLett.104.026801,2010NatNa...5..703D,2007NatPh...3..455Y} The same holds true for molecular electronics, where the primary goal is the understanding of quantum transport through individual nanoscale objects such as molecules, carbon nanotubes or DNA.\cite{Nitzan30052003,2000Natur.407...57P,2002Natur.419..906S,2002Natur.417..722P} In this case quantum dots can have intrinsic vibrational degrees of freedom, or local phonons. Dealing with superconductor hybrids with local phonons is desirable but complicated\cite{PhysRevB.73.214501,Bai2011661,PhysRevB.81.104508,PhysRevB.81.104508} due to the manifold energy-scales present in the system. So far the focus of most theoretical studies has either been on a SC hybrid in the presence of Coulomb interaction\cite{PhysRevB.63.094515,PhysRevB.84.075484,PhysRevB.84.115448,PhysRevB.84.155403,PhysRevB.68.035105,PhysRevLett.99.126602} or normal conducting systems in the presence of a phonon mode.\cite{scheer98,PhysRevLett.100.176403,2011arXiv1104.4903K,PhysRevB.82.165116,0953-8984-23-10-105301,PhysRevB.86.081412,PhysRevB.87.085127,alvaro2008,carmina2009,carmina2010}
Until now, exact theoretical treatments of such many-body systems in the presence of a phonon mode or Coulomb interaction have been put forward only for normal conducting systems.\cite{PhysRevLett.100.176403,PhysRevB.85.075118,PhysRevB.81.165106,PhysRevB.81.035108,PhysRevLett.97.076405,PhysRevLett.101.140601,PhysRevLett.107.206801,PhysRevB.83.165315,wang:244506} Time-dependent phenomena in superconducting systems have only been analyzed in the long-time limit \cite{PhysRevB.85.075301} or the adiabatic limit\cite{PhysRevB.84.155403} yet, but a complete description of the case of a 'preparative' nonequilibrium, i.e. the time evolution of the system from some initial preparation towards its steady state, under finite external bias voltage so far has again only been investigated in the case of normal conducting systems.\cite{2012arXiv1207.6222J,PhysRevB.80.125109,PhysRevB.87.085127,PhysRevB.78.235110,PhysRevB.79.245102,PhysRevB.50.5528,PhysRevB.71.165321,0953-8984-19-37-376206,PhysRevB.74.085324,PhysRevB.86.081412}\\
In this paper we attempt to provide first steps towards the understanding of
the transient and steady state behavior of the Anderson-Holstein model with a
superconducting lead. In Section \ref{model} of this paper we describe the
model used for the SC-quantum dot-normal (SQN) junction and how it can be
mapped to an effective model in the limit of strong electrostatic interactions
but no electron-phonon interaction. Section \ref{analytical} is devoted to a
description of an analytical approach for the model without phonon
interaction. An approximative approach based on a a rate equation
description\cite{2011arXiv1104.4903K} is proposed to treat phononic couplings
to the dot in Section \ref{rate}. The numerically exact diagrammatic Monte Carlo (diagMC) simulation method is briefly discussed in Section \ref{monte1}. Finally, in Section \ref{results} the results from the analytical approach and the rate equation description are compared with diagMC data.
\section{Model} \label{model}
The full Hamiltonian of the Anderson-Holstein model with a superconducting lead consists of seven contributions

\begin{align}
  H = H_{\mathrm{dot}} + H_{\text{n}} + H_{\text{s}} +  H_{\text{T}} + H_{\text{U}} + H_{\mathrm{dot,Ph}}^{(I)} + H_{\mathrm{Ph}}
  \,. 
  \label{hsys}
\end{align}

\begin{figure}
  \includegraphics[width=8cm]{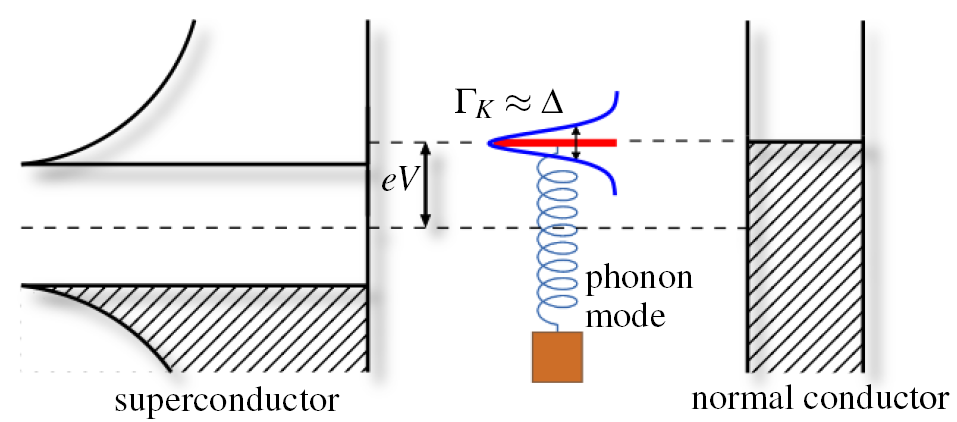}
  \caption{Sketch of the considered system: a quantum dot in the regime of strong onsite interaction is assumed to be in the Kondo regime so that its resonance width is given by $\Gamma_{\text{K}}$. It is coupled to a normal conductor and a SC where we have indicated the specific energy-dependence of the DOS on the superconducting side. The quantum dot is subjected to a finite voltage bias coupled to a phonon mode.}
  \label{fig1}
\end{figure}

The situation considered in this work is also sketched in Fig.~\ref{fig1}.

The spin-degenerate quantum dot is given by a single electronic level at the energy $\Delta_{\text{d}}$

\begin{eqnarray}
  H_{\mathrm{dot}} = \sum_\sigma \Delta_{\text{d}} d_\sigma^{\dagger} d_\sigma
  \,,
\end{eqnarray}

where we use units such that $e = \hbar = k_{\text{B}} = 1$. The normal electrode $n$ is described as a fermionic continuum in terms of field operators $R_\sigma(x)$ at chemical potential $\mu_{\text{R}}$. For the SC we use a standard BCS Hamiltonian with its characteristic energy gap $\Delta$ using field operators $L_\sigma(x)$

\begin{eqnarray}
  H_{\text{s}} &=& \sum_{k,\sigma} \epsilon_k L_{k\sigma}^{\dagger} L_{k\sigma} \nonumber\\
  && + \Delta \sum_k (L_{k\uparrow}^{\dagger} L_{-k\downarrow}^{\dagger} + L_{-k\downarrow} L_{k\uparrow})
  \,.
\end{eqnarray}

While the density of states (DOS) on the normal conductor is assumed to be constant for all energies contributing to the electron transport, $\rho_{\text{R}}(\omega)=\rho_{0\text{R}}$, the superconducting lead has an energy-dependent DOS 

\begin{align}
  \rho_{\text{L}}(\omega) 
  &= 
  \frac{
    \rho_{0 \text{L}} 
    |\omega|
  }{
    \sqrt{\omega^2 - \Delta^2} 
  }
  \theta
  \left(
    \frac{
      |\omega|-\Delta
    }{
      \Delta
    }
  \right)
  \label{eq:superconducting_DOS}
  \,.
\end{align}

As in previous treatments of SC hybrids\cite{PhysRevB.50.3982,0295-5075-91-4-47004,PhysRevB.76.184510} we define the voltage with respect to the chemical potential of the superconducting lead $\mu_{\text{L}} = 0$ so that $V= - \mu_{\text{R}}$. Tunneling is assumed to occur locally between the dot and the leads

\begin{eqnarray}
  H_{\text{T}} &=& \sum_{\alpha = \text{L,R}} \sum_\sigma \gamma_\alpha [\alpha_\sigma^{\dagger}(x=0) d_\sigma + \mathrm{h.c.}]
  \,,
\end{eqnarray}

where $\gamma_{\alpha}$ refers to the tunneling amplitude from lead $\alpha=$ L,R to the dot. Additionally $H_{\text{U}}$ accounts for the electron-electron onsite interaction of the system

\begin{eqnarray}
  H_{\text{U}} = U (d_\uparrow^{\dagger} d_\uparrow -1/2) (d_\downarrow^{\dagger} d_\downarrow - 1/2)
  \,,
\end{eqnarray}

where $U$ denotes the strength of the Coulomb repulsion. The final ingredient is the electron-phonon interaction

\begin{eqnarray}
  H_{\mathrm{dot,Ph}}^{(I)} = \sum_\sigma d_\sigma^{\dagger} d_\sigma \lambda_0 (b^{\dagger} + b)
  \,,
  \label{eq:hamiltonian_electron_phonon_interaction}
\end{eqnarray}

which couples the electronic degrees of freedom on the dot to a local phonon mode described by

\begin{eqnarray}
  H_{\mathrm{Ph}} = \omega_0 (b^{\dagger} b + 1/2)
  \,.
  \label{eq:hamiltonian_phonons}
\end{eqnarray}

The general solution of the system described by Eq.~(\ref{hsys}) seems to be out of question and approximations (at least in order to proceed analytically) are thus necessary. In the first part we therefore consider the case $\lambda_0 = 0$, meaning without electron-phonon interaction. We want to consider the case of strong onsite interaction and an odd number of electrons on the quantum dot. In this case the most prominent effect is the emergence of a spin 1/2 Kondo resonance around the Fermi edge\cite{0957-4484-15-7-056} for temperatures below the Kondo temperature $T_{\text{K}}$. In normal conducting systems the Kondo temperature is directly related to the onsite interaction via\cite{doi:10.1080/00018738300101581}

\begin{eqnarray}
  T_{\text{K}} = \frac{\sqrt{2 U \Gamma_n}}{\pi} \exp \left(-\pi U /8\Gamma_n\right)
  \,,
\end{eqnarray}

where $\Gamma_n$ refers to the tunnel rate between the normal
conductor and the quantum dot. Due to the additional energy scale $\Delta$ in
our problem, two scenarios may occur: for large $T_{\text{K}}/\Delta$ the Kondo
resonance couples to the quasiparticles in the SC leading to a behavior similar
to the one for normal conducting
systems\cite{0957-4484-15-7-056,PhysRevB.63.094515} whereas for small
$T_{\text{K}}/\Delta$ the Kondo resonance is weakly coupled to the SC due to
the absence of mobile electrons at the Fermi edge.\cite{PhysRevLett.89.256801}
This transition as a function of the onsite interaction has been discussed in
detail in Ref.~[\onlinecite{PhysRevB.63.094515}]. In the present work we investigate the latter case since it shows clearer signatures of superconducting correlations.\\
In this regime one can greatly simplify the problem. Below the gap electronic
transport in SC hybrids is due to Andreev
reflection.\cite{doi:10.1080/00018732.2011.624266} However, in presence of strong onsite interaction and weak coupling of the SC to the quantum dot a double occupancy of the latter is very unlikely and Andreev processes are strongly suppressed.\cite{Soller2011425,PhysRevB.55.R6137} An effective model in this case is the resonant level model, where the role of the hybridization with the normal lead is given by $T_{\text{K}}$.\cite{springerlink:10.1007/BF00654541,0957-4484-15-7-056,Soller2011425} Due to the suppression of Andreev reflection from now on we treat the superconductor as a normal conductor with an energy-dependent DOS given by $\rho_{\text{L}}(\omega)$.\cite{PhysRevB.55.R6137} Studies of new experiments on the Kondo resonance in SC hybrids\cite{PhysRevLett.104.246804} revealed that the background density of states can be very small compared to the Kondo peak\cite{soller3} so that the system may indeed be effectively described by a resonant level between the SC and the normal conductor with strongly asymmetric hybridizations of the quantum dot with the SC and the normal lead for voltages not much larger than $T_{\text{K}}$. The Kondo resonance is then pinned to the Fermi level in the normal lead so that $\Delta_{\text{d}} = \mu_{\text{R}}$.\\
Nonetheless, the Kondo effect is a collective phenomenon and facing the case of
a `preparative' nonequilibrium one could argue that such a mapping is not
adequate since the time scale for the central Kondo peak to fully develop
should be given by $1/T_{\text{K}}$.\cite{PhysRevLett.83.808} Recent studies of
the transient behavior of the Anderson impurity model, however, revealed that
the central peak develops much faster.\cite{PhysRevB.83.075107}
Therefore the model considered should not only be a specific case of the
resonant level model but also a good approximate description of
the behavior in the deep Kondo limit.\cite{2012arXiv1205.0876O}

\section{Analytical approach} \label{analytical}

We first describe the analytical approach to the transient dynamics of the SQN
junction using the mapping to a resonant level model described above. Since
for the resonant level model expectation values separate into spin-up and
spin-down contributions we work with spinless operators from now on. As in Ref.~[\onlinecite{PhysRevB.78.235110}] we have to deal with two different initial
situations: (i) the dot level is empty ($n_0=0$) and (ii) the dot
is populated by one electron ($n_0 = 1$) which, however, lead to
the same steady-state current, so that only (i) is investigated. Due to the
simple structure of $H_{\mathrm{dot}}$ the time evolution of the uncoupled dot
is trivial and we obtain the following Keldysh Green's function (GF)

\begin{widetext}
  \begin{eqnarray}
    \mathbf{D}_0(t) = \left[\begin{array}{cc} D_0(t) & D_0^<(t) \\ D_0^>(t) & \tilde{D}_0(t)\end{array}\right] = e^{- i\Delta_{\text{d}} t} \left[\begin{array}{cc} - i [\theta(t) ( 1-n_0) - \theta(-t) n_0] & i n_0\\ - i(1- n_0) & - i [\theta(-t) (1-n_0) - \theta(t) n_0] \end{array}\right]
    \,,
  \end{eqnarray}
\end{widetext}

where $D_0(t)$ and $\tilde{D}_0(t)$ refer to the time-ordered and anti-time-ordered GFs, respectively. For the retarded and advanced components we have

\begin{eqnarray}
  D_0^{\text{r}}(t) &=& D_0(t) - D_0^<(t) = - i \theta(t) e^{- i \Delta_{\text{d}} t}\,,\\
  D_0^{\text{a}}(t) &=& D_0^<(t) - \tilde{D}_0(t)  = i \theta(-t) e^{- i \Delta_{\text{d}} t}\,.
\end{eqnarray}

Compared to Ref.~[\onlinecite{PhysRevB.78.235110}] we now have to deal with two
different lead GFs. Mind that due to the suppression of Andreev reflection the anomalous GFs are zero. In the normal and superconducting case the retarded
GF has the form

\begin{eqnarray}
  g_{\alpha}^{\text{r}}(t) = - i \theta(t) \int \text{d} \omega \rho_\alpha(\omega) e^{- i \omega t}\,.
\end{eqnarray}

The DOS of the superconductor, $\rho_{\text{L}}(\omega)$, depends on energy due to the superconducting correlations [see Eq.~(\ref{eq:superconducting_DOS})]. $\rho_{\text{R}}(\omega)$ can also become energy-dependent due to a finite bandwidth as discussed in [\onlinecite{PhysRevB.78.235110}]. While for our analytical approach we consider the case of a wide flat band $\rho_{\text{R}}(\omega) = \rho_{0\text{R}}$ an energy cutoff has to be introduced to employ the diagMC simulation method in Section \ref{monte1}. One obtains the full Keldysh matrix

\begin{eqnarray}
  \mathbf{g}_\alpha(\omega) = i 2\pi \rho_\alpha(\omega) \left[\begin{array}{cc} n_\alpha - \frac{1}{2} & n_\alpha \\ -(1-n_\alpha) & n_\alpha - \frac{1}{2}\end{array}\right]\,,
\end{eqnarray}

where $n_\alpha$ denotes the Fermi distribution function in the respective electrode. The retarded and advanced components are given by $g_\alpha^{\text{r}}(\omega) = - i \pi \rho_\alpha(\omega)$ and $g_\alpha^{\text{a}}(\omega) = [g_\alpha^{\text{r}}(\omega)]^*$. The GFs of the coupled system can now be found for any time dependence of $\gamma_\alpha(t)$ but in this Section we concentrate on the case of sudden switching where $\gamma_\alpha(t) = \gamma_\alpha \theta(t)$. The time-evolution of the retarded GF is given by the standard expression\cite{PhysRevB.50.5528,0022-3719-4-8-018}

\begin{align}
  D^{\text{r}}(t,t') &= D_0^{\text{r}}(t-t') + \int_0^\infty \text{d} t_2 K(t,t_2) D^{\text{r}}(t_2,t')
  \,, 
  \label{fulld}
\end{align}

where

\begin{eqnarray}
  K(t,t_2) &=& K_{\text{R}}(t,t_2) + K_{\text{L}}(t,t_2)\\
  &=& \int_0^\infty \text{d} t_1 D_0^{\text{r}}(t-t_1) \Sigma_{\text{R}}^{\text{r}}(t_1 - t_2) \nonumber\\
  && + \int_0^\infty \text{d} t_1 D_0^{\text{r}}(t-t_1) \Sigma_{\text{L}}^{\text{r}}(t_1 - t_2)\,,
\end{eqnarray}

is the kernel involving the lead self-energies

\begin{eqnarray}
  \Sigma_{\alpha}^{\text{r}}(t) = \gamma_\alpha^2 g_\alpha^{\text{r}}(t)\,,
\end{eqnarray}

The lead self-energy $\boldsymbol{\Sigma} = \boldsymbol{\Sigma}_{\text{L}} + \boldsymbol{\Sigma}_{\text{R}}$ is given by

\begin{eqnarray}
  \Sigma_{\text{L}}^{\text{r}}(t) &=& - \frac{ i \theta(t)}{4\pi} \int_{-\infty}^\infty \text{d} \epsilon \frac{e^{- i \epsilon t} \Gamma_{\text{L}} |\epsilon|}{\sqrt{\epsilon^2 - \Delta^2}} \theta\left(\frac{|\epsilon| - \Delta}{\Delta}\right)\,, \label{sigmal}\\
  \Sigma_{\text{R}}^{\text{r}}(t) &=& - \frac{ i \theta(t)}{4\pi} \int_{-\infty}^\infty \text{d} \epsilon \; e^{- i \epsilon t} \Gamma_{\text{R}}\,, \label{sigmar}
\end{eqnarray}

with $\Gamma_{\text{L}} = 2 \pi \rho_{0 \text{L}} \gamma_{\text{L}}^2$ and $\Gamma_{\text{R}} = 2 \pi \rho_{0 {\text{R}}} \gamma_{\text{R}}^2$.\\
From the calculation for the normal conducting case we know that the integral equation for the retarded GF involving the normal lead

\begin{align}
  D_{\text{R}}^{\text{r}}(t,t') = D_0^{\text{r}}(t-t') + \int_0^\infty \text{d} t_2 K_{\text{R}}(t,t_2) D_{\text{R}}^{\text{r}}(t_2,t')\,,
\end{align}

can be solved by iterations\cite{PhysRevB.43.2541,PhysRevB.50.5528,PhysRevB.78.235110} leading to

\begin{eqnarray}
  D_{\text{R}}^{\text{r}}(t-t') = - i \theta(t-t') e^{- i \Delta_{\text{d}} (t-t')} e^{- \Gamma_{\text{R}}/2 (t-t')}\,. \label{dnormal}
\end{eqnarray}

For the superconducting part we first do the Fourier transformation in Eq.~(\ref{sigmal}) following Ref.~[\onlinecite{PhysRevB.30.6419}]

\begin{eqnarray}
  \int_0^\infty \text{d} \epsilon \frac{|\epsilon|}{\sqrt{\epsilon^2 - \Delta^2}} e^{- i \epsilon t} \theta\left(\frac{|\epsilon| - \Delta}{\Delta}\right) = K_1 (\Delta t)\,,
\end{eqnarray}

where $K_1$ is the modified Bessel function [\onlinecite{stegun}]. Using this result we arrive at

\begin{eqnarray*}
  K_{\text{L}}(t,t') = - \frac{\theta(t-t')}{\pi} \int_{t'}^t \text{d} t_1 e^{- i \Delta_{\text{d}} (t-t_1)}\ K_1(\Delta(t_1 - t'))\,,
\end{eqnarray*}

which we have to evaluate numerically in the expression for the full retarded GF in Eq.~(\ref{fulld}). Due to the numerical evaluation of $K_{\text{L}}(t,t')$ we cannot solve this Dyson equation analytically. However, using Eq.~(\ref{dnormal}) as the starting point for the iterations in Eq.~(\ref{fulld}) and following our considerations in Section \ref{model} we know that $\Gamma_{\text{L}} \ll \Gamma_{\text{R}}$ and two iterations of $D^{\text{r}}$ are sufficient and can be performed numerically.\\
A simply accessible quantity for the resonant level model is the time-dependent dot population $n(t) = \langle d^{\dagger}(t) d(t)\rangle$ which can be directly expressed via the off-diagonal Keldysh GF

\begin{eqnarray}
  n(t) = - i D^<(t,t)\,.
\end{eqnarray}

\begin{figure}
  \includegraphics[width=8cm]{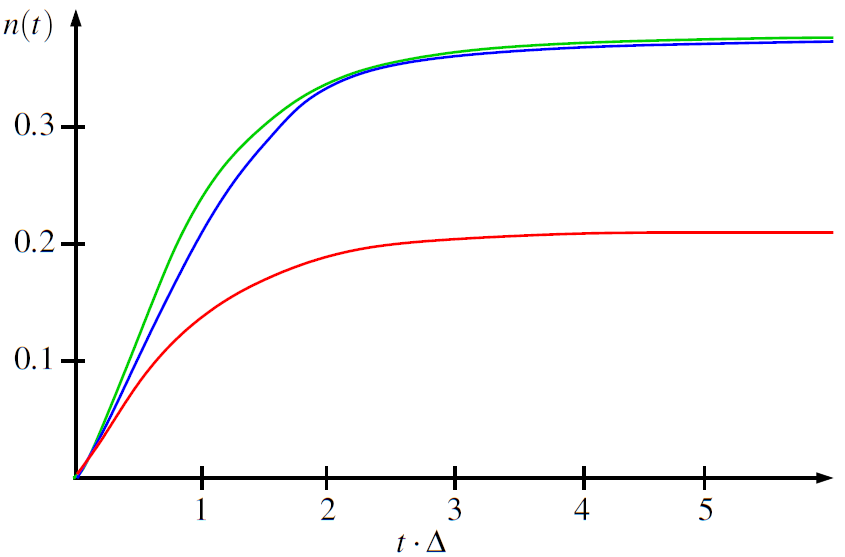}
  \caption{Time-dependent dot population $n(t)$. The graph shows the result for $T=0.1 \Delta, \; \Gamma_{\text{R}} = \Delta, \; \Gamma_{\text{L}} = 0.05 \Delta$ and $\Delta_{\text{d}} = -V$. The lowest (red) curve refers to $V=0$, the middle (blue) curve to $V=\Delta$ and the upper (green) curve is for $V=2\Delta$. The decisive time scale is $1/\Gamma_R$.}
  \label{fig2}
\end{figure}

The lesser GF can be expressed by the already known retarded GF via\cite{PhysRevB.78.235110}

\begin{eqnarray}
  D^< = (1+ D^{\text{r}} \Sigma^{\text{r}}) D_0^< (1+ \Sigma^{\text{a}} D^{\text{a}}) + D^{\text{r}} \Sigma^< D^{\text{a}}\,,
\end{eqnarray}

where  in the product notation integrations over inferred variables are implied and $\Sigma = \Sigma_{\text{L}} + \Sigma_{\text{R}}$ is defined. This relation is especially suited for the case of an initially empty dot since then $D_0^< = 0$ and only the last term contributes.\\
Typical experiments\cite{PhysRevLett.104.246804,0957-4484-15-7-056} have $T_{\text{K}} = \Gamma_{\text{R}} \lesssim \Delta$. Therefore, one can consider $\Gamma_{\text{R}} = 0.9 \Delta$ and $\Gamma_{\text{L}} = 0.1 \Delta$ in order to stay within the limits of applicability of our approach. Typical results for the time-dependent dot population are shown in Fig. \ref{fig2}. Of course, we refer to the dot population of the specific resonant level model treated here and not to the one of a full Kondo system.
\\
Compared to the results for two normal conducting leads coupled to a quantum dot \cite{PhysRevB.78.235110} we observe no strong population oscillations due to the strongly asymmetric couplings of the dot to the leads.\\
The more important observable is the current through the system. We can rewrite the current on the normal conducting side in terms of Keldysh GFs and perform the Keldysh disentanglement as in Ref.~[\onlinecite{PhysRevB.78.235110}]

\begin{eqnarray}
  I_{\text{R}}(t) &=& I'_{\text{R}}(t) + I''_{\text{R}}(t)\,,\\
  I'_{\text{R}}(t) &=& - \gamma_{\text{R}}^2 \mathrm{Re} \int_0^\infty \text{d} t_1 g_{\text{R}}^{\text{r}}(t,t_1) D^{\text{K}}(t_1,t)\,,\\ 
  I''_{\text{R}}(t) &=& \gamma_{\text{R}}^2 \mathrm{Re} \int_0^\infty \text{d} t_1 D^{\text{r}}(t,t_1) g_{\text{R}}^{\text{K}}(t_1, t)\,.
\end{eqnarray}

The first contribution is essentially given by the time-dependent dot population

\begin{eqnarray}
  I'_{\text{R}}(t) = \frac{\Gamma_{\text{R}}}{2} \theta(t) [1-2n(t)]\,.
\end{eqnarray}

As $I_R''(t)$ starts at zero, as in the case of normal contacts, there is an instantaneous current onset $I_{\text{R}}(0) = \Gamma_{\text{R}}/2$ which stems from the fact that we assume a sudden switching and have taken the wide band limit so that electrons at arbitrarily high energies are able to occupy the dot in correspondingly fast processes.\\
Access to the current on the left side can either be obtained via the GFs directly but this way is cumbersome since the calculation of the current on the normal side is easier due to the simpler structure of the GFs. Instead we can use the displacement current $I_{\mathrm{disp}}$ which is given by the time derivative of $n(t)$\cite{PhysRevLett.70.4114,PhysRevB.66.195319}

\begin{eqnarray}
  I_{\mathrm{disp}}(t) = \frac{\text{d} n(t)}{\text{d} t}\,.
\end{eqnarray}

It represents the difference in currents between the right and left side

\begin{eqnarray}
  I_{\mathrm{disp}}(t) &=& I_{\text{R}}(t) - I_{\text{L}}(t)\,,
\end{eqnarray}

so that

\begin{eqnarray}
  I_{\text{L}}(t) &=& I_{\text{R}}(t) - I_{\mathrm{disp}}(t)\,.
\end{eqnarray}

Consequently $I_{\text{L}}(t)$ will show a similar current onset as $I_{\text{R}}$ since calculating $D^{\text{r}}$ in Eq. (\ref{fulld}) with two iterations in $\Gamma_{\text{L}}$ regularizes the initial derivative of $n(t)$. We may also define the average current through the junction as

\begin{eqnarray}
  I(t) = [I_{\text{R}}(t) + I_{\text{L}}(t)]/2\,. \label{average}
\end{eqnarray}

Including an electron-phonon coupling the current is accessed by means of an approximative rate equation description discussed in the next Section as well as the numerically exact Monte Carlo method outlined in Section \ref{monte1}.

\section{Rate Equation Approach to Electron-phonon Coupling} \label{rate}

The next step is to include the electron-phonon coupling described by Eq.~(\ref{eq:hamiltonian_electron_phonon_interaction}). The mapping to a resonant level in the deep Kondo limit, described in Section \ref{model} including onsite interaction, is not expected to hold in the presence of arbitrary electron-phonon coupling. Still, at moderate electron-phonon interaction strength it has been shown in [\onlinecite{PhysRevLett.94.176801}] that the Kondo effect does not disappear at once but rather persists, however, associated with a change in $T_{\text{K}}$ and the appearance of phonon sidebands. Therefore, in the case of moderate electron-phonon interaction strengths the mapping to an effective Kondo model is still possible and a good qualitative description could be obtained from a rate equation approach.\cite{PhysRevLett.76.1715} In this situation we have a resonant level model coupled to phonons as in [\onlinecite{2011arXiv1104.4903K}], however, with a SC density of states. We note that for the resonant level model with normal conducting leads more involved analytical calculations have been performed either perturbatively [\onlinecite{PhysRevB.73.075428,PhysRevB.72.201101,PhysRevB.80.041307,PhysRevB.77.113405,PhysRevLett.103.136601,PhysRevB.80.041309,PhysRevB.82.121414}] or in a nonperturbative way [\onlinecite{PhysRevB.83.085401}].\\
For the resonant level model it is convenient to work with dressed electronic
states by applying a polaron transformation $U_p = \exp[(\lambda_0/\omega_0)
d^{\dagger} d (b^{\dagger} - b)]$ which leads to a Hamiltonian where the electron-phonon interaction $H_{\mathrm{dot,Ph}}^{(I)}$ is completely absorbed in the tunnel part of the Hamiltonian meaning that we are left with Eq. (\ref{hsys}) in the form of $H_{\mathrm{sys}} = H_{\text{L}} + H_{\text{R}} + \tilde{H}_{\text{T}} +  H_{\mathrm{Ph}} + H_{\mathrm{dot}}$, where

\begin{eqnarray}
  \tilde{H}_{\text{T}} &=& \sum_{\alpha = L,R} \gamma_\alpha [\alpha^{\dagger}(x=0) e^{(\lambda_0/\omega_0) (b^{\dagger} -b)} d  + \mathrm{h.c.}]\,, \nonumber\\
  \tilde{H}_{\mathrm{dot}} &=& (\Delta_{\text{d}} + \lambda_0^2/\omega_0) d^{\dagger} d\,. \label{pol2}
\end{eqnarray}

We absorb the polaron shift of the dot energy by a redefinition of $\Delta_{\text{d}} \rightarrow \Delta_{\text{d}} - \lambda_0^2/\omega_0$. Again using the self-energies introduced in Eqs. (\ref{sigmal}) and (\ref{sigmar}) we obtain the forward and backward rates onto the dot ($\Gamma_1, \; \Gamma'_2$) and away from the dot ($\Gamma'_1, \; \Gamma_2$) following [\onlinecite{2011arXiv1104.4903K}]

\begin{eqnarray*}
  \Gamma_1(V) &=& \int \frac{\text{d} \epsilon}{2\pi} \frac{\Gamma_{\text{L}} |\epsilon|}{\sqrt{\omega^2 - \Delta^2}} \theta \left(\frac{|\epsilon| -  \Delta}{\Delta}\right) n_{\text{L}} P(\epsilon - V)\,,\\
  \Gamma_2(V) &=& \int \frac{\text{d} \epsilon}{2\pi} \Gamma_{\text{R}} (1- n_{\text{R}}) P(\epsilon + V)\,,\\
  \Gamma'_1(V) &=& \int \frac{\text{d} \epsilon}{2\pi} \frac{\Gamma_{\text{L}} |\epsilon|}{\sqrt{\omega^2 - \Delta^2}} \theta \left(\frac{|\epsilon| -  \Delta}{\Delta}\right) (1- n_{\text{L}}) P(\epsilon + V)\,,\\
  \Gamma'_2(V) &=& \int \frac{\text{d} \epsilon}{2\pi} \Gamma_{\text{R}} n_{\text{R}} P(\epsilon - V)\,.
\end{eqnarray*}

The inelastic tunneling processes associated with the emission and absorption of phonons are described by the function $P(\epsilon)$ being the Fourier transform of the phonon-phonon correlation function. For this correlation function we assume that the phonons are thermally distributed, which may be due to coupling to a thermal environment given by the substrate or a backgate.\cite{PhysRevB.69.245302} The effect of coupling to an external bath can be characterized by an additional coupling constant $\gamma_{\text{B}}$, and in the following we assume the bath to be purely ohmic. In this case the phonon spectral density has Lorentzian shape\cite{RevModPhys.59.1}

\begin{eqnarray}
  J(\epsilon) &=& \frac{\gamma_{\text{B}} \omega}{[(\epsilon/\omega_0)^2 -1]^2 + [\gamma_{\text{B}} \omega_0 \epsilon /(2\lambda_0^2)]^2}\,.
\end{eqnarray}

The phonon-phonon correlation function can now be calculated analytically leading to

\begin{widetext}
  \begin{eqnarray}
    P(\epsilon) = \frac{e^{- \rho_{\gamma_{\text{B}}}}}{\pi} \mathrm{Re} \left\{\sum_{k,l
        = 0}^\infty \frac{\rho_{\gamma_{\text{B}},a}^k}{k!}
      \frac{\rho_{\gamma_{\text{B}}, e}^l}{l!} \frac{i}{\epsilon+ \Omega_0 k - \Omega_0^*
        l + i \Gamma_{\mathrm{tot}}/2} \right\}\,,
  \end{eqnarray}
  
  where $\Omega = \omega_0 \xi + i \gamma_{\text{B}} /2, \; \xi = \sqrt{1-
    \gamma_{\text{B}}^2 /(4\omega_0^2)}$ and $\Gamma_{\mathrm{tot}} = \Gamma_{\text{L}} + \Gamma_{\text{R}}$. The functions $\rho_{\gamma_{\text{B}}}$, $\rho_{\gamma_{\text{B}}, \alpha}$ and $\rho_{\gamma_{\text{B}}, e}$ are given by
  
  \begin{eqnarray}
    \rho_{\gamma_{\text{B}}} &=& \frac{\lambda_0^2}{2 \omega_0 \sqrt{\omega_0^2 - \gamma^2/4}} \left[\frac{\coth\left(\frac{\beta \Omega}{2}\right)}{\Omega^2} + \frac{\coth \left(\frac{\beta \Omega^*}{2}\right)}{(\Omega^*)^2}\right]\,,\\
    \rho_{\gamma_{\text{B}},a} &=& \frac{\lambda_0^2}{2 \omega_0 \sqrt{\omega_0^2 - \gamma_{\text{B}}^2/4}}\frac{\coth\left(\frac{\beta \Omega}{2}\right) -1}{\Omega^2}, \; \rho_{\gamma_{\text{B}},e} = \frac{\lambda_0^2}{2 \omega_0 \sqrt{\omega_0^2 - \gamma_{\text{B}}^2/4}}\frac{\coth \left(\frac{\beta \Omega^*}{2}\right)-1}{(\Omega^*)^2}\,.
  \end{eqnarray}
\end{widetext}

Using the above defined rates in the master equation for the dot population we can solve for the steady state and derive the current which is given by

\begin{eqnarray}
  I(V) &=& \frac{\Gamma_1 \Gamma_2 - \Gamma'_1 \Gamma'_2}{\Gamma_1 + \Gamma_2 + \Gamma'_1 + \Gamma'_2}\,. \label{currentpe}
\end{eqnarray}

We have checked that a slight coupling to the environment $\gamma_B$ only leads to minor quantitative changes of our results so that we will treat the case $\gamma_{\text{B}} = 0$ in the following.\\
For voltages of the order of $\omega_0$ effects of the electron-phonon interaction are most pronounced. Therefore, we will focus on this region in order to calculate the behavior of the current depending on the voltage. Further, the voltage is increased from values smaller than the superconducting gap to ones which are larger. This behavior is then compared to the case of a pure metallic two terminal setup in order to investigate the effect of the superconductor in the system.

\section{Diagrammatic Monte Carlo method} \label{monte1}

Using a diagrammatic expansion in terms of the tunneling coupling, time-dependent observables of the nonequilibrium Anderson-Holstein model can be conveniently simulated numerically exact by means of Monte Carlo methods. \cite{PhysRevLett.100.176403,PhysRevB.83.075107,Werner2009} In this paper, we focus on the calculation of the time-dependent current from lead $\alpha$ to the dot, which can be written as a sum over time-ordered integrals \cite{PhysRevLett.100.176403}

\begin{align}
  I_{\alpha}(t) 
  &=
  2
  \sum \limits_{n = 1}^{\infty}
  (-1)^{n}
  \int \limits_0^{t}
  \text{d} \vec{s}_n
  \ 
  \text{Re} 
  \left \lbrace
    \mathcal{L}_{\alpha}(\vec{s}_n)
    \mathcal{G}(\vec{s}_n)
  \right \rbrace
  \label{eq:diagrammatic_expansion}
  \,.
\end{align}

We used the abbreviation

\begin{align}
  \int \limits_0^{t}
  \text{d} \vec{s}_n
  \equiv
  \int \limits_0^{t}
  \text{d}s_{n}
  \int \limits_0^{s_{n}}
  \text{d}s_{n-1}
  \cdots
  \int \limits_0^{s_2}  
  \text{d}s_1
  \,,
\end{align}

where $\vec{s} = \{ s_1,s_2,\dots,s_{n} \}$ is the time-ordered sequence of $n$ tunneling times $s_j$ where an electron tunnels from a lead to the quantum dot or vice versa. 

Initially a factorizing preparation is assumed, where the leads are decoupled from the quantum dot. The coupling of the leads to the dot is performed at $t=0$ either instantaneously, or smoothly within some switch-on time $\tau_{\text{sw}}$ using a sine function (for details see Ref.~[\onlinecite{PhysRevB.83.075107}]). 

The influence of the leads degrees of freedom are given by a determinant of a matrix consisting of the uncoupled leads' lesser and greater self energies

\begin{align}
  \mathcal{L}_{\alpha} (\vec{s}) 
  &= 
  i^n 
  \det 
  \left \lbrace 
    \mathcal{S}^{(\alpha)}(\vec{s})
  \right \rbrace
  \label{eq:leads_influence}
  \,,
\end{align}

where again no anomalous contributions need to be considered due to the suppression of Andreev reflection. The matrix $\mathcal{S}^{(\alpha)}$ is given by

\begin{align}
  \mathcal{S}^{\alpha}_{j,k}(\vec{s})
  &=
  \left \lbrace
    \begin{matrix} 
      \Sigma^{<}(s_{2k-1},s_{2j}) \mbox{, for } j \le k
      \\ 
      \Sigma^{>}(s_{2k-1},s_{2j}) \mbox{, for } j > k
    \end{matrix}
  \right.
  \label{eq:leads_determinant}
  \,,
\end{align}

by replacing $\Sigma^{</>}$ with $\Sigma^{</>}_{\alpha}$ whenever one time argument is equal to the final propagation time $t$. The lesser and greater self-energies are obtained using the DOS of the leads truncated at the finite energy $\pm \epsilon_{\text{c}}$ so that $\rho_{\alpha}(\omega)=\rho_{0\alpha} \theta(\epsilon_{\text{c}}-|\omega|)$. Such a cutoff might lead to small deviations of the transient behavior when compared with results for the wideband limit. For the long-time behavior, however, these discrepancies vanish.\cite{PhysRevB.78.235110}

The self-energies of the normal conductor including a cutoff can be calculated analytically.\cite{PhysRevB.78.235110} For the case of the superconducting lead a numerical evaluation is necessary.

The influence of the phonons and the energy level of the dot are described via

\begin{align}
  \mathcal{G}(\vec{s})
  &=
  \mathcal{F}[\vec{s}]
  \
  e^{
    i 
    \Delta_{\text{d}} 
    \left(
      s_1
      -
      s_2
      +
      \cdots
    \right)
  }
  \label{eq:phonon_influence}
  \,,
\end{align}

where ${\mathcal F}[\vec{s}_n]$ denotes the Feynman-Vernon influence functional\cite{1963AnPhy..24..118F} given by

\begin{align} 
  {\mathcal F}[\vec{s}_n] 
  &= 
  \exp 
  \left \lbrace 
    -
    \int \limits_{\mathcal{C}} 
    \text{d}s_1 \!\!\!\!\!
    \int \limits_{\mathcal{C}: s_2 < s_1} \!\!\!\!\!
    \text{d}s_2 q(s_1) L(s_1-s_2) q(s_2) 
  \right \rbrace 
  \label{eq:feynman_vernon}
  \,.
\end{align}

The integrations are performed on the Keldysh time contour $\mathcal{C}: 0 \to t \to 0$.  The second integration over $s_2$ has to fulfill $s_2 < s_1$ on this contour.  $q(s)$ denotes the occupation of the quantum dot at time $s$ for a given set of tunneling times. Thus, it is fully determined by the initial condition of the quantum dot -- empty or occupied -- and the times as well as number of the electron tunneling times given by $\vec{s}_n$. For the considered single phonon mode, the bath autocorrelation function, used to calculate Eq.~(\ref{eq:feynman_vernon}), is

\begin{align} 
  L(s) 
  &= 
  \frac{\lambda_0^2}{\omega_0} 
  \left[ 
    \cos(\omega_0 s) 
    - 
    i
    \sin(\omega_0 s) 
  \right] 
  \,.
\end{align}

Using Eq.~(\ref{eq:leads_influence}) and Eq.~(\ref{eq:phonon_influence}) the influence of the leads and the dot on the current in Eq.~(\ref{eq:diagrammatic_expansion}), given by $\text{Re} \left \lbrace \mathcal{L}_{\alpha}(\vec{s}_n) \mathcal{G}(\vec{s}_n) \right \rbrace$, can be readily calculated numerically for a given sequence of tunneling times $\vec{s}_n$ without any approximation. The remaining task to calculate the current which corresponds to the summation and integration over all possible tunneling events in Eq.~(\ref{eq:diagrammatic_expansion}).  This can be done conveniently in a numerically exact manner employing Monte Carlo techniques.\cite{PhysRevLett.100.176403,Werner2009} Using this method, the only occurring error is a statistical one.

Despite being numerically exact, the diagMC has a drawback since it is suffering from the dynamical sign problem: the stochastic Monte Carlo sum has to be performed over terms with alternating signs. This leads to small average values for the observable with large statistical errors so that the CPU time scales exponentially with the system time. Nevertheless, times of the order of $10\Delta^{-1}$ can be accessed within a reasonable simulation time so that the long-time behavior of the current can be addressed.

\section{Results}\label{results}

\subsection{SQN junction without electron-phonon interaction}

\begin{figure}[h!]
  \includegraphics[width=8cm]{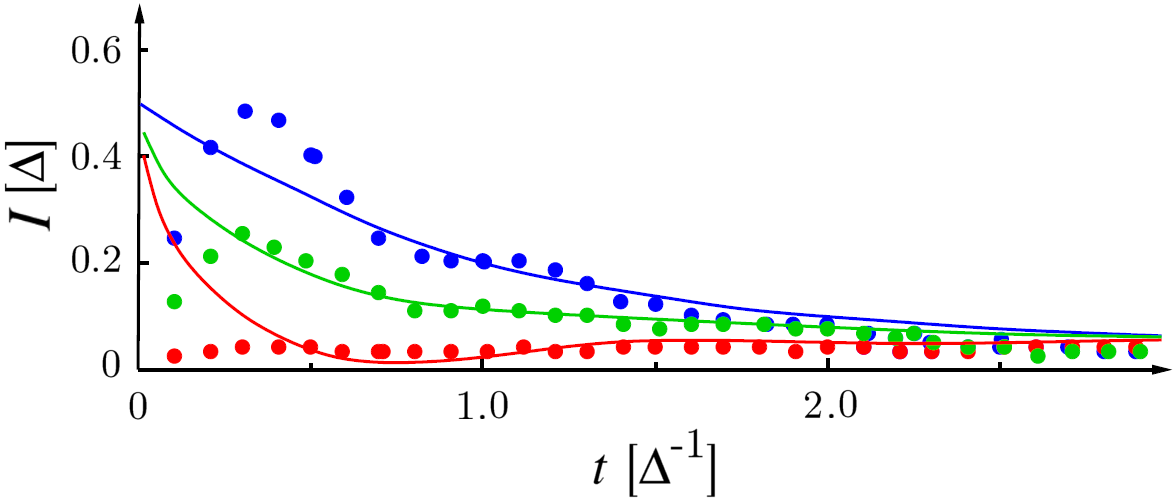}
  \caption{Time-dependent current through the quantum dot for an instantaneous
    switch-on at $t=0$. We choose the parameters $V= 2 \Delta, \Delta_{\text{d}}
    = -2 \Delta, \; \Gamma_{\text{R}} = \Delta, \; \Gamma_{\text{L}} = 0.05
    \Delta$ and $T = 1/\beta =0.1 \Delta$. The diagMC results are calculated with
    a bandwidth of $2\epsilon_{\text{c}} = 20 \Delta$ whereas for the analytical calculation we use the wideband limit. The right current is
    highlighted with blue color, the left one is red and the average one green,
    respectively. Straight lines denote the analytical results, the dots the
    diagMC ones.} 
  \label{fig3}
\end{figure}

In this Section the SQN junction without electron-phonon interaction is studied in order to discuss the effects caused by the superconductor in comparison to a pure metallic two terminal setup. The time-dependent dynamics are analyzed by comparing the analytical approach of Section \ref{analytical} to the numerically exact results from the Monte Carlo method. This provides an accurate test for our analytical approach and further reveals details of the influence of the SC gap on the electron transport in the absence of phonon coupling.

We use parameters $\Gamma_{\text{R}} = \Delta, \; \Gamma_{\text{L}} = 0.05 \Delta, \; \beta = 10 \Delta$ with a voltage $V= 2 \Delta$ larger than the superconducting gap. A comparison of the analytical approach with the Monte Carlo method for time dependent left, right and average current is shown in Fig. \ref{fig3}.

For small times we observe a deviation of the diagMC results from the
analytical approximation, which can be explained by different bandwidths in the
two cases. As already discussed for normal conducting systems such differences
in the bandwidths lead to deviations for small times.\cite{PhysRevB.78.235110} For times $t \gtrsim \Delta^{-1}$ the agreement is good for all currents, and a steady state value is reached at around $t \gtrsim 2.5 \Delta^{-1}$.

\begin{figure}[h!]
  \includegraphics[width=8cm]{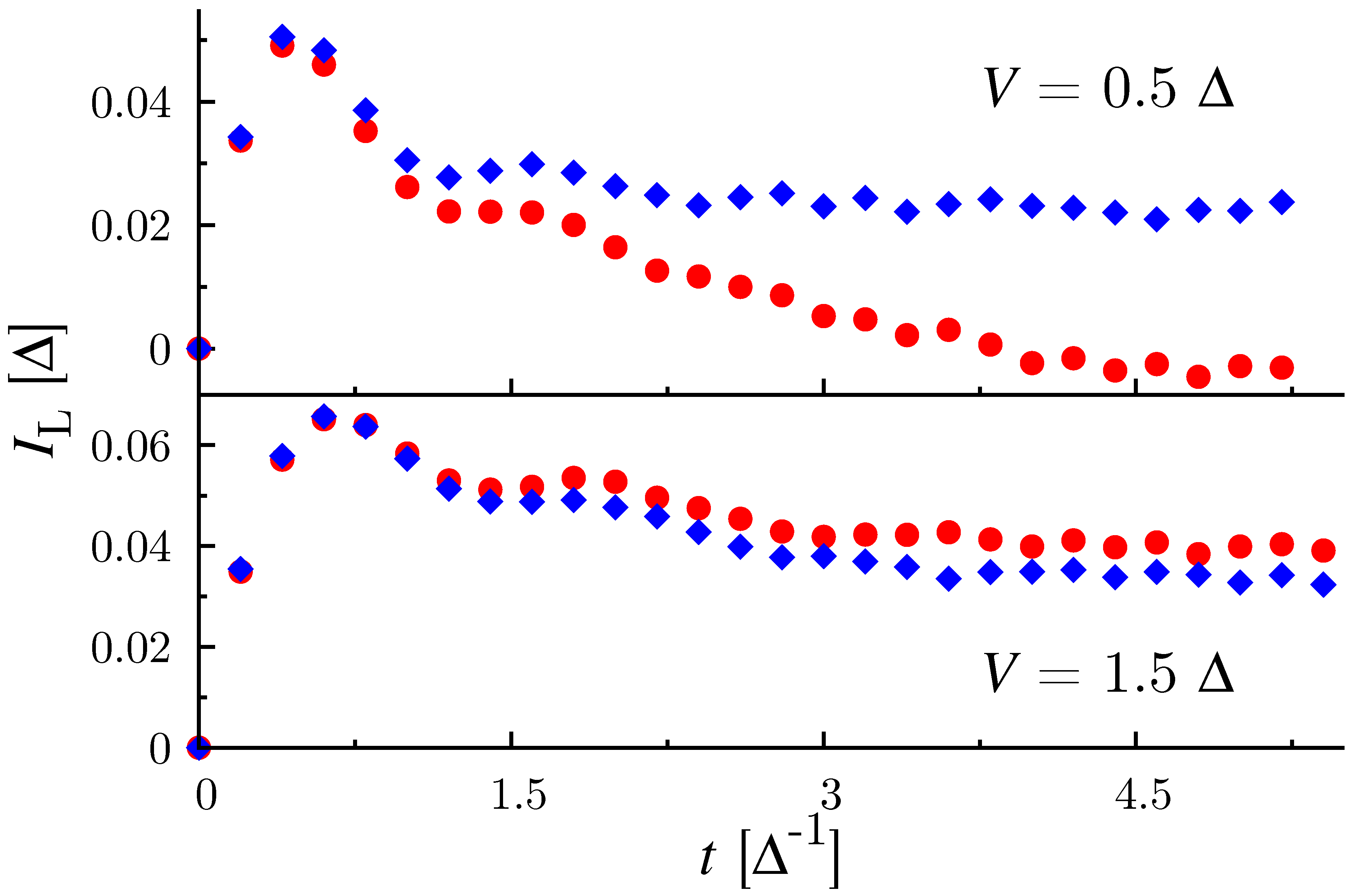}
  \caption{Comparison of the time-dependent currents for a pure metallic setup (blue) with a SQN junction (red) for a voltage smaller than the superconducting gap $V=0.5\Delta$ (upper panel) and larger $V=1.5\Delta$ (lower panel). The other parameters are $\Gamma_{\text{R}} = 0.9 \Delta, \; \Gamma_{\text{L}} = 0.1\Delta,$ and $T = 1/\beta =0.1 \Delta$. For the Monte Carlo results (symbols: diamonds (metallic) and filled circles (SQN)) a bandwidth of $2\epsilon_{\text{c}} = 12 \Delta$ is used whereas for the analytical calculations are performed in the wideband limit. Note that $\Delta_{\text{d}}=-V$ for our model (see Section \ref{model}).} 
  \label{Fig:v0p5_v1p5_noninteracting}
\end{figure}

In accordance with previous studies for normal
contacts\cite{PhysRevLett.100.176403,PhysRevB.78.235110} we observe a fast
approach to the steady state even in the situation of strongly asymmetric
couplings. The observed time scale in this situation is given by $1/
\Gamma_{\text{R}}$, which allows for reliable and fast simulation of the system. Indeed, the time-dependent behavior is very similar to normal conducting systems since it is mainly controlled by the large initial current onset on the normal conducting side due to the strongly asymmetric hybridization. On the other hand, for voltages well below the superconducting gap, the current from a pure metallic setup is quite different to the one of an SQN junction as depicted in the to panel of Fig.~\ref{Fig:v0p5_v1p5_noninteracting}.

Right after the coupling from the dot to the leads, the currents for the two different setups show a similar transient. The reason for this similar behavior is that for a quick coupling of the leads to the dot high excitations contribute to the transient \cite{PhysRevB.78.235110} so that the superconducting gap plays a minor role immediately after the coupling. For larger times, however, a steady state is established for the metallic setup at about $t \geq 2 \Delta^{-1}$ whereas the current for the SQN setup remains transient and finally reaches a plateau at around $t \geq 4 \Delta^{-1}$. While the extracted steady-state current for the metallic setup shows a finite value, the current for the SQN junction is vanishing since the superconducting gap is too large to be overcome.

For a voltage $V=1.5\Delta$ the transients of the two setups reveal a very similar behavior, however, the current for the SQN junction is always slightly larger than the metallic one. Note that this difference vanishes when further increasing the voltage (not shown).

The effect of the superconductor on the electron transport through the dot is studied including an electron-phonon interaction in the next Sections.

\subsection{Time-dependent dynamics with a vibrational mode}

As a first step for discussing the SQN setup with a vibrational mode, the transient current is calculated and compared with a metallic two terminal setup with a single phonon mode. We would like to stress that while the time-dependent dynamics can be described with the diagMC method the rate equation approach provides no direct access to them. Therefore, in this Section we calculate the time dynamics of the system via diagMC and then compare the extracted steady-state current with the results from the rate equation approach in the next Section.

The basic setup is the same as in the previous Section but additionally the quantum dot is coupled to a single phonon mode of frequency $\omega_0 = 2 \Delta$ for a moderate interaction strength $\lambda_0 = 2 \Delta$. 

\begin{figure}[h!]
  \includegraphics[width=8cm]{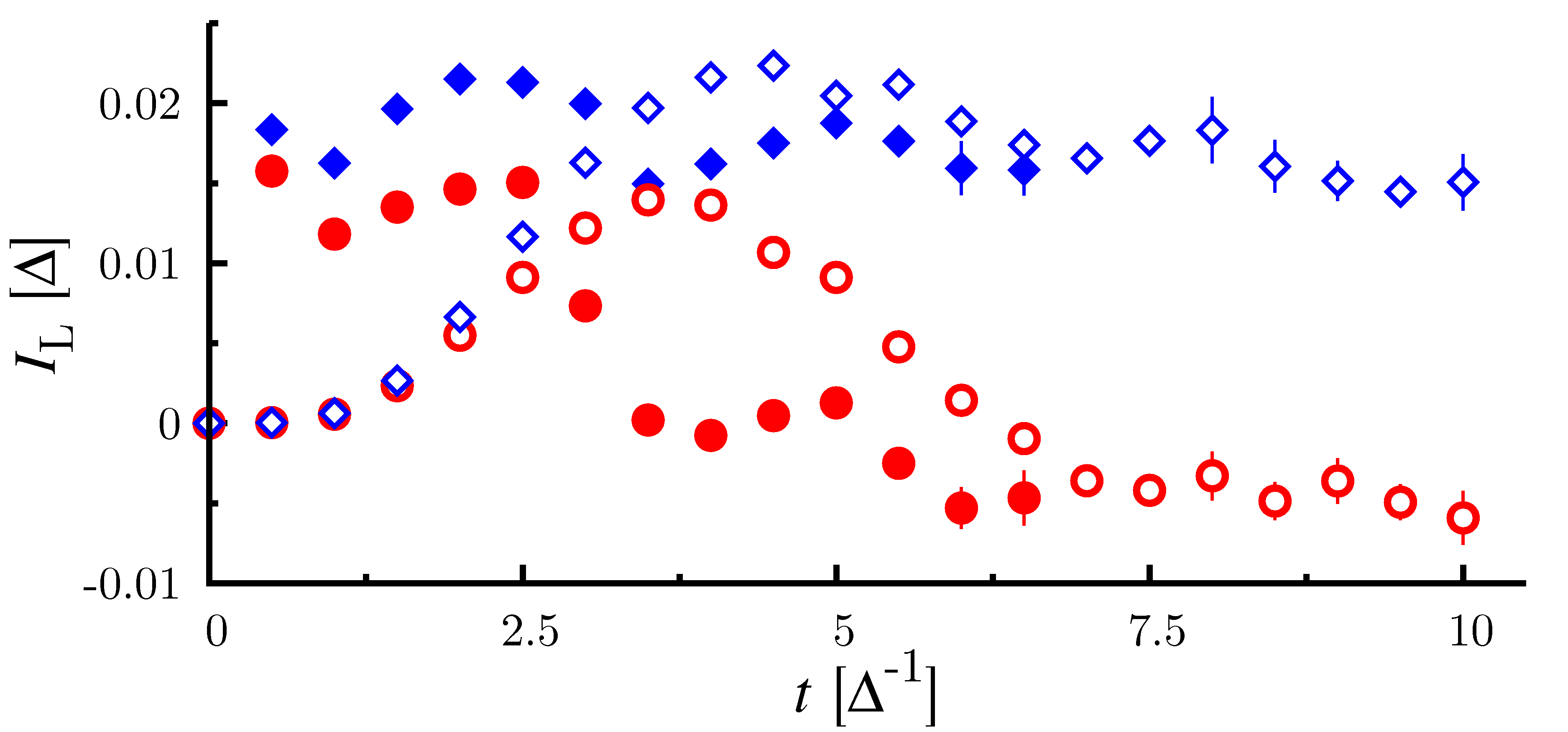}
  \caption{Time-dependent currents for a setup with two metallic leads (blue) and a SQN junction (red) for a voltage smaller than the superconducting gap $V=0.5\Delta=-\Delta_{\text{d}}$ and the parameters as in Fig.~\ref{Fig:v0p5_v1p5_noninteracting} including a phonon mode with $\lambda_0=\omega_0=2\Delta$. An instantaneous switching of the leads to the dot is denoted by filled symbols, whereas a smooth switching within the switch-on time $\tau_{\text{sw}}=4\Delta^{-1}$ is depicted with empty symbols.} 
  \label{Fig:current_left_v0p5}
\end{figure}

The time-dependent tunneling current between the superconducting lead and the quantum dot for a voltage smaller than the size of the superconducting gap $V=0.5\Delta$ is shown in Fig.~\ref{Fig:current_left_v0p5}. The diagMC results of the SQN junction are compared with the ones for a setup with two metallic leads. Two different switching methods of the coupling between the leads and the dot at $t=0$ are depicted: an instantaneous and a smooth one.

Compared with the case of no electron-phonon interaction (see top panel of Fig.~\ref{Fig:v0p5_v1p5_noninteracting}) the vibrational mode has a severe influence on the transient behavior. The time scales for reaching the steady state are similar, however, an oscillatory behavior stemming from the phonon is clearly visible for an instantaneous switching. The phononic degree of freedom is excited by the high initial current. After this first ``phonon shake-up'', however, these high energy excitations no longer contribute to the electron transport through the quantum dot so that a drop in the current is observed along with typical oscillations associated with the phonon frequency. This leads to a step-like decrease of the current for the SQN setup instead of an (almost) linear one for the noninteracting case. These oscillations vanish in the case of a smooth switch-on procedure where one does not start from a strongly nonequilibrium situation as in the sudden-switching case. 

The steady-state current for the metallic setup is smaller than for the case of no electron-phonon coupling. For the SQN junction the long-time current drops to a small, but finite and negative value. From the DOS it is expected that in the long time limit this current will vanish. Thus for the times accessible by diagMC still phonon oscillations around the steady state are still observable.

\begin{figure}[h!]
  \includegraphics[width=8cm]{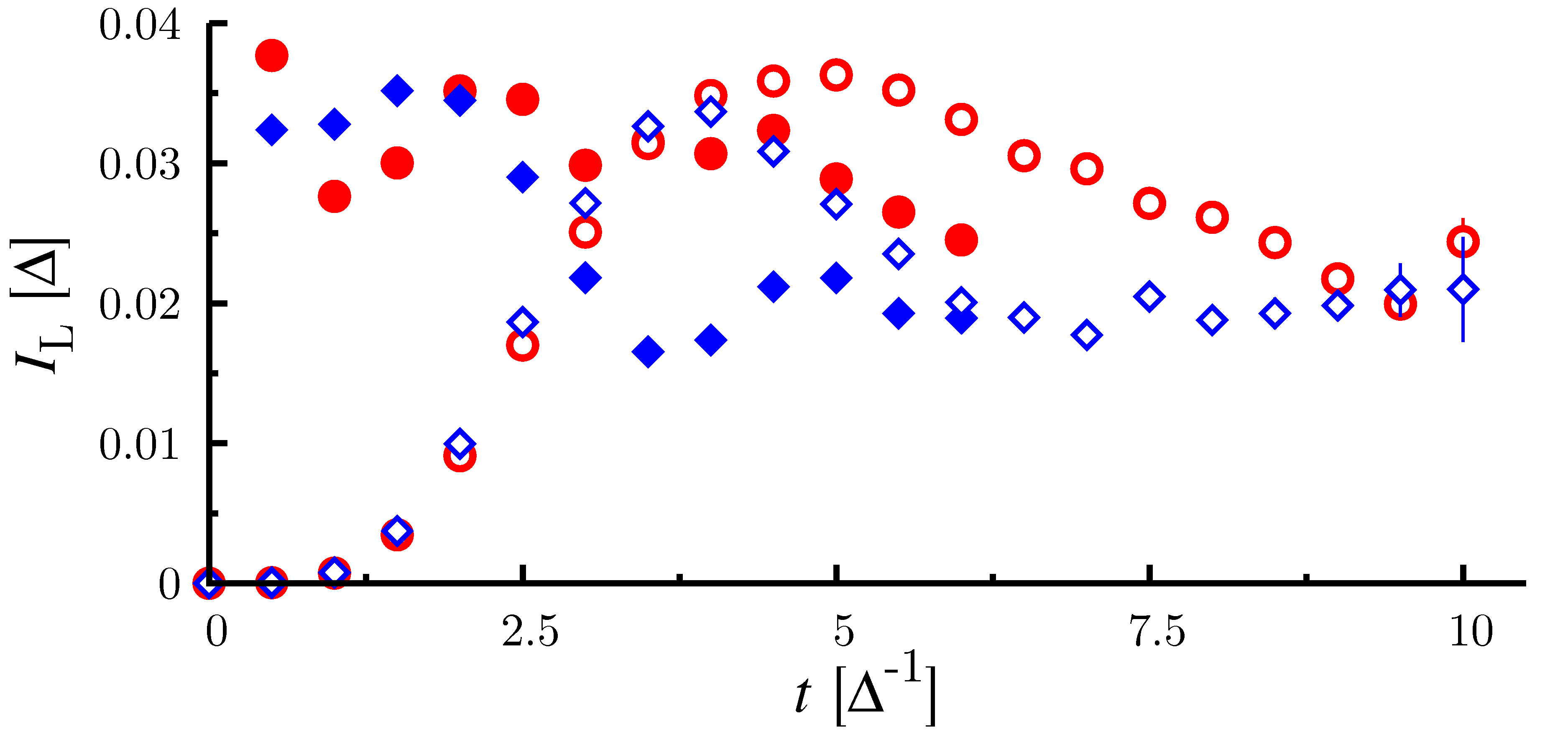}
  \caption{Same color code and parameters as in Fig.~\ref{Fig:current_left_v0p5} but with a voltage above the superconducting gap $V=1.5\Delta=-\Delta_{\text{d}}$. The phonon influence is too strong to extract a long-time current for the SQN setup.}\label{Fig:current_left_v1p5}
\end{figure}

Increasing the voltage to be larger than the superconducting gap $V=1.5\Delta$ the voltage is sufficiently high so that transport channels with energies larger than the superconducting gap can be also used in the SQN setup. Therefore, the long time current is also finite. The transient of the SQN junction is larger than the one of the metallic setup for most times as presented in Fig.~\ref{Fig:current_left_v1p5}. For an instantaneous switching, the current reveals an overall behavior similar to the case without phonons, however, with significant phonon oscillations. Here, the current of the SQN junction is larger than the one for the metallic setup for values larger than $t \gtrsim 2.5\Delta^{-1}$. From the instantaneous switching no steady state can be extracted due to the phonon oscillations. Note that the period for the phonon oscillations is independent of the applied bias voltage.

Using a smooth switching of the leads to the dot the long-time current of the metallic setup is accessible for $t \gtrsim 6 \Delta^{-1}$ where only small phonon oscillations around the steady state remain. For the SQN setups a smooth switching reduces the oscillation but the influence of the phonons on the transient is still strong. This can be seen by the large overshooting of the transient current for the smooth switching. The transient currents of the SQN setups for the instantaneous and smooth switching hint for a larger long-time current, which provide a sign for more pronounced phonon sidebands in the SQN spectral function compared with the metallic one. However, it is not possible to confirm this behavior with the Monte Carlo method since the time scales accessible are too small here. In order to confirm this finding we have to compare our simulations with a rate equation method which allows to directly extract the steady-state current as presented in Section ~\ref{steady_state_results}.

\begin{figure}[h!]
  \includegraphics[width=8cm]{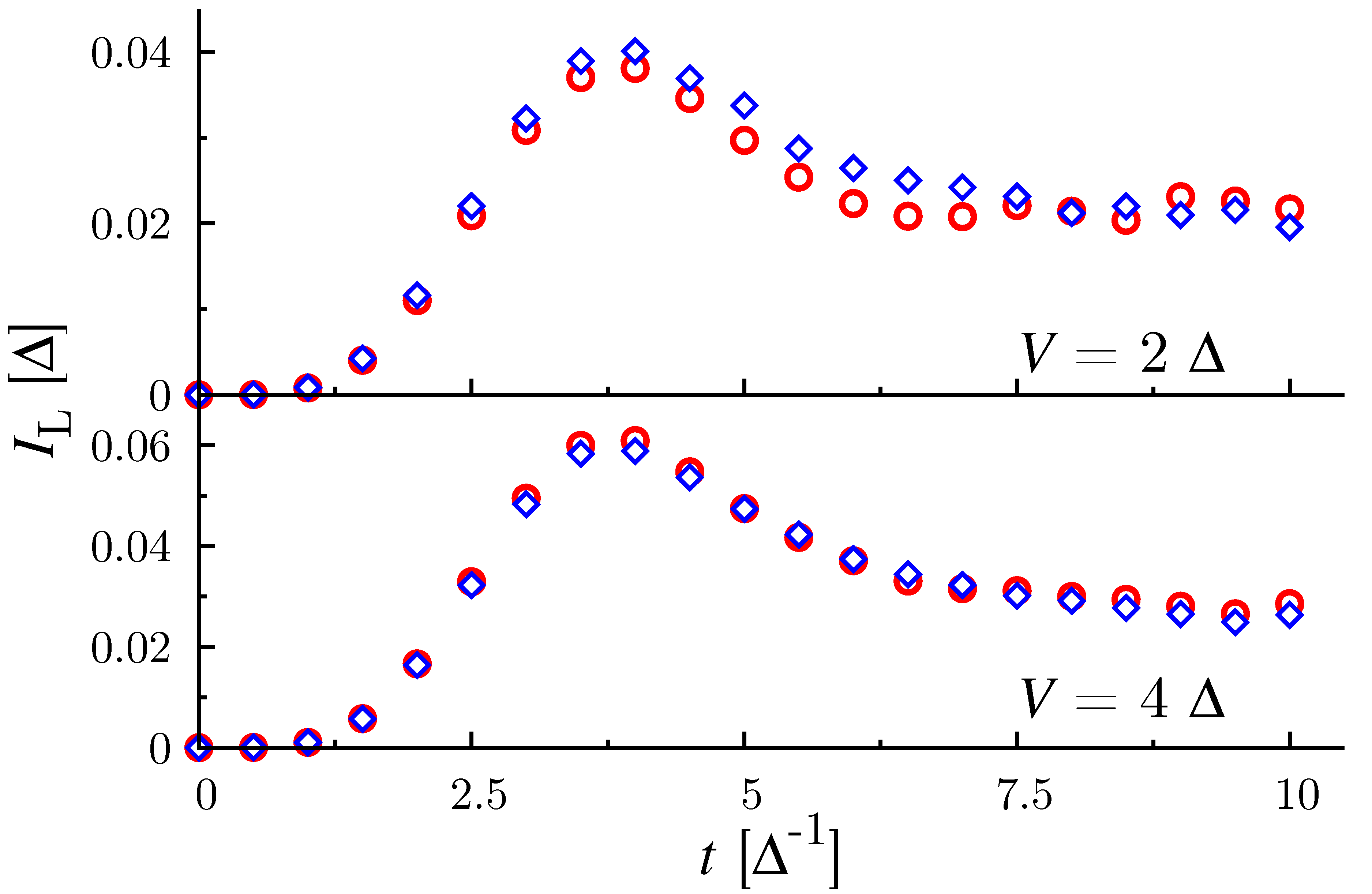}
  \caption{Time-dependent current for $V = 2 \Delta$ and $V= 4\Delta$ for the same parameters and color code as in Fig.~\ref{Fig:current_left_v0p5}.
    \label{Fig:current_left_v2p0_v4p0}}
\end{figure}

For voltages well above the superconducting gap, the transients and long-time current are similar for the SQN and the metallic setup as shown in Fig.~\ref{Fig:current_left_v2p0_v4p0}. Two voltages, $V=2\Delta$ and $V=4\Delta$ are accessible for a smooth switching of the leads to the dot.

The transient show an initial overshooting and then a smooth convergence towards their steady state which can then be extracted and compared with the rate equation approach in the next Section.

\subsection{Steady state results}\label{steady_state_results}

\begin{figure}
  \includegraphics[width=8cm]{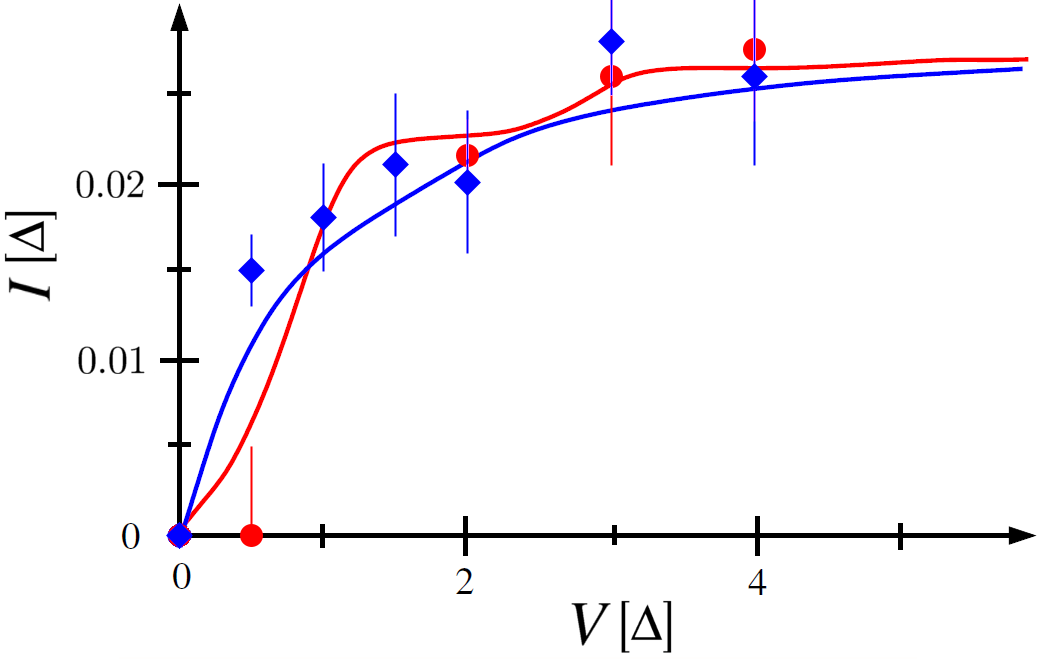}
  \caption{Steady state current as a function of voltage given by the rate
    equation in Eq. (\ref{currentpe}) (solid curves) and the diagMC approach
    (dots). The result is for a moderate coupling $\lambda_0 = 2 \Delta$, $\omega_0 = 2
    \Delta$ for the phonon mode and $\beta = 10 \Delta$. The red curve and
    the red dots correspond to the superconducting case (rate equation and
    diagMC, respectively) and the blue curve corresponds to the normal conducting
    case using the same $\Gamma_L, \; \Gamma_R$.}
  \label{fig5}
\end{figure}

The results for the $I-V$ curve are shown in Fig.~\ref{fig5} from both the rate equation and diagMC. It is easy to see that both methods agree reasonably well within the statistical errors of the diagMC data.

The pronounced step like feature in the data for the SC is due to the interplay of the phonon and the SC DOS. The convolution of the two leads to steps in the current not at multiples of the oscillator frequency but equally spaced by $V \approx \omega_0$.

\begin{figure}
  \includegraphics[width=8cm]{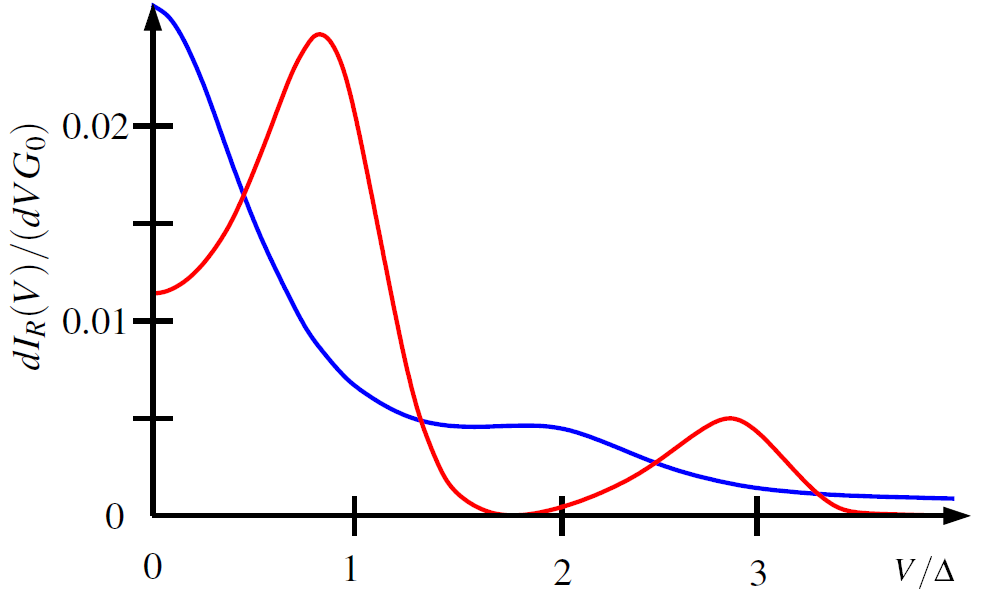}
  \caption{Conductance $dI_{\text{R}}(V)/dV$ using the same parameters as for the plot in Fig7. \ref{fig5}. The comparison of the conductance in the superconducting (red) and normal (blue) case shows that the conductance peaks due to the phonon sidebands become much more pronounced.}
  \label{fig6}
\end{figure}

These effects are even more pronounced in the conductance $dI_{\text{R}}(V)/dV$. Since the rate equation approach provides us with a continuous curve we can calculate the derivative shown in Fig. \ref{fig6} and compare it to the normal conducting case.

The conductance shows that the steps become much more pronounced compared to the normal conducting case due to the SC DOS. Thus we expect that even in the deep Kondo limit where the effective model is also of the resonant level model type, the sidebands will be enhanced as soon as one of the leads becomes superconducting.\\
Finally, we want to comment on the possibility of experimental confirmation of our findings. Typical superconductor hybrid structures\cite{2010NatNa...5..703D} using Al as the superconductor have $\Gamma \approx T_K \approx 100$meV and the superconductor is only weakly coupled.\cite{0957-4484-15-7-056} A typical value for short carbon nanotubes or molecular junctions\cite{2002Natur.417..725L,2009NatPh...5..327L} is $\omega_0 \approx 100-1000$meV. Measurements of the steady-state current can then be done straightforwardly. Transient current spectroscopy\cite{PhysRevB.63.081304} is more demanding but also seems within reach.

\section{Conclusion}

In conclusion, we have investigated the transient dynamics and the steady-state behavior of the Anderson-Holstein model with a superconducting lead. In absence of a phonon mode, we investigated the transient dynamics of the model with a superconducting lead using a mapping of the deep Kondo limit to the resonant level model. We have found good agreement between an analytical and a diagMC approach. Including an electron-phonon onsite interaction the transient current was studied in detail employing the diagMC method. For small voltages the current of the setup strongly depends on the SC gap, whereas for large voltages the results are similar to a metallic two terminal setup. For voltages of the order of the phonon frequency we found signs for more pronounced phonon sidebands. This effect was studied in detail using a rate equation approach and again observed good agreement for the steady state current. Electron-phonon interaction leads to more pronounced sidebands in the spectral function compared to the normal conducting case which can be useful for nanoscale transistors. Our results are both interesting for future experimental applications and also provide a benchmark for future investigations.\cite{PhysRevLett.90.246403,PhysRevLett.96.216802,PhysRevB.77.033409}

The authors would like to thank D. Kast, J. Ankerhold, A. Levy Yeyati, S. Maier, K. Joho and L. Hofstetter for many interesting discussions. KFA acknowledges the computational resources provided by the bwGRID. AK is supported by CQD of the University of Heidelberg.

\end{document}